\documentclass[a4paper,10pt]{article}
\usepackage[english]{babel}

\usepackage{authblk}
\usepackage{xcolor}
\usepackage{graphicx} 
\usepackage{amsmath}
\usepackage{biblatex} 
\newcommand{\W}[2]{%
\mathbf{W}_\text{#1}^\text{(#2)}}

\newcommand{\Wrc}[1]{%
\mathbf{W}_\text{#1}}
\newcommand{\X}[2]{%
\mathbf{x}_{#1}^\text{(#2)}}

\newcommand{\Xrc}[1]{%
\mathbf{x}_{#1}}
\newcommand{\Y}[2]{%
{y}_{#1}^{(#2)}}
\newcommand{\Yrc}[1]{%
{y}_{#1}}

\newcommand{\frc}[1]{%
f_\text{#1}}
\newcommand{\usig}[2]{%
u_{#1}^{(#2)}}
\newcommand{\usigrc}[1]{%
u_{#1}}

\newcommand{\sidecaption}[1]
{\raisebox{\abovecaptionskip}{\begin{subfigure}[c]{1.6em}
  \caption[singlelinecheck=off]{}
  \label{#1}
\end{subfigure}}\ignorespaces}

\usepackage{subcaption}

\title{Deep Photonic Reservoir Computer Based on Frequency Multiplexing with Fully Analog Connection Between Layers}

\author[1, *]{Alessandro Lupo}
\author[1]{Enrico Picco}
\author[2]{Marina Zajnulina}
\author[1]{Serge Massar}
\affil[1]{Laboratoire d'Information Quantique, Université libre de Bruxelles (ULB), Bruxelles, Belgium.}
\affil[2]{Multitel Innovation Centre, Mons, Belgium}
\affil[*]{alessandro.lupo@ulb.be}

\begin{document}

\maketitle

\begin{abstract} 
Reservoir computers (RC) are randomized recurrent neural networks well adapted to process time series, performing tasks such as nonlinear distortion compensation or prediction of chaotic dynamics. Deep reservoir computers (deep-RC), in which the output of one reservoir is used as \textcolor{black}{the input for another one}, can lead to improved performance because, as in other deep artificial neural networks, the successive layers represent the data in more and more abstract ways. 
We present a fiber-based photonic implementation of a two-layer deep-RC based on frequency multiplexing. The two RC layers are encoded in two frequency combs propagating in the same experimental setup. The connection between \textcolor{black}{the} layers is fully analog and does not require any digital processing. We find that the deep-RC outperforms \textcolor{black}{a} traditional RC by up to two orders of magnitude on two benchmark tasks. This work paves the way towards using fully analog photonic neuromorphic computing for complex processing of time series, while avoiding costly analog-to-digital and digital-to-analog conversions.
\end{abstract}

\section{Introduction}

Artificial Intelligence is probably the most disruptive new technology to emerge during the first decades of the XXI century. Its success is based on the use of deep neural networks in which multiple layers of artificial neurons are connected in a feed\textcolor{black}{-}forward architecture \cite{Bengio2009,LeCun2015}.   
Recent advances include, for instance, image classification and analysis \cite{ard2022five}, game playing \cite{silver2018general}, protein structure prediction\cite{doi:10.1126/science.abj8754,jumper2021highly}, chatbots that simulate human conversation such as ChatGPT and Bing\cite{chatgpt, bing}, and more.

Artificial neural networks are fundamentally analog systems simulated on a digital computer. Thus, it seems \textcolor{black}{highly} attractive to replace the digital simulation \textcolor{black}{with} analog hardware, as this could result in considerable energy savings. Photonics is particularly attractive for analog \textcolor{black}{implementation of neural networks due to} its potential for very ultra speed (see e.g.\ \cite{xu202111,feldmann2021parallel}), parallelism (see e.g.\  \cite{Liutkus2014,SaadeRandomProj2016}), possibility of implementing spiking networks (see e.g.\ \cite{feldmann2019all,jha2022photonic}), and low energy consumption per operation (see e.g.\ \cite{hamerly2019large}).
The importance of deep neural networks for complex applications has \textcolor{black}{led} to several demonstrations of deep photonic networks \textcolor{black}{using on-chip integrated optics \cite{shen2017deep,ashtiani2022chip}, optoelectronics \cite{zhou2021large}, and a 3D-printed stack of diffractive surfaces \cite{lin2018all}. These configurations replicate the mathematical concept of artificial neural networks, i.e.\ they physically implement layers of linear and nonlinear transformations of the input by deploying optical and opto-electronic components.} 

\textcolor{black}{However, in order to realize analog implementations of artificial neural networks, one should try to exploit as much as possible the natural dynamics of the employed physical system. Some neural network algorithms, such as extreme learning machines (ELM) \cite{huang2006extreme} or reservoir computers (RC) \cite{Jaeger04, Tanaka2019, wetzstein2020inference, markovic2020physics}, are more amenable to physical implementations because only a part of their weights is trained, while the rest can be chosen at random. These random connections can often be replaced by the inherent dynamics of the physical system without loss of performance.}
 
\textcolor{black}{Reservoir computers, which are the topic of the present work, are randomized recurrent neural networks (RNN) in which the recurrence is provided by a (simulated or physical) high-dimensional nonlinear dynamical system called a “reservoir" \cite{Jaeger04}. RCs have been successfully implemented in analog systems including photonics, electronics, spintronics, mechanics, biology, and more (see \cite{Tanaka2019} for a review). Many photonic RC implementations use a delay loop and a single dynamical node to encode multiple neurons by means of time multiplexing (as proposed in \cite{Appeltant11}), see e.g.\ \cite{brunner2013parallel,Larger17}.
Although the time multiplexing architecture is simple to implement, it suffers from an intrinsic slowdown because the time to process an input will be given by the number of neurons multiplied by the time to process a single neuron. Alternative approaches that do not suffer from this inherent slowdown use a form of parallelism such as spatial multiplexing (in free space optics \cite{Rafayelyan2020} or multimode fibers \cite{Sunada2020}), a hybrid temporal/spatial approach \cite{nakajima2021scalable}, or frequency multiplexing \cite{Butschek2022}.}

\textcolor{black}{As in the case with other types of neural networks,} assembling several RCs in a deep architecture enhances \textcolor{black}{the overall system performance in data processing}. Deep RCs were first used in \cite{triefenbach2010phoneme} and studied in more depth in \cite{Gallicchio2017},
where it is shown that the serial connection among different RC layers enhances the system performance by enriching its dynamics. Different ways of combining \textcolor{black}{(in series or in parallel)} photonic reservoirs in\textcolor{black}{to} networks are compared in \cite{freiberger2019improving}. Motivated by these works, \textcolor{black}{the} first experimental implementation of \textcolor{black}{a} deep-RC is reported in \cite{Nakajima2022}, showing significant improvement in performance \textcolor{black}{when the number of layers is increased}. However, in this work each reservoir was implemented using the time multiplexing architecture, which is not optimal in terms of computing speed, and, more importantly, the connection between reservoirs was implemented digitally. The latter is also the case in the related work \cite{wright2022deep}. Ref.\ \cite{lin2022deep} proposes an architecture for a deep reservoir based on time delay architecture with analog connection between \textcolor{black}{the} layers.
 
Here\textcolor{black}{, we report a deep reservoir configuration consisting of two interconnected reservoir layers with a fully analog connection that does not require data storage or processing on a digital computer.} Our experiment is based on a recently reported \textcolor{black}{RC} in which the neuron signals are encoded in the amplitudes of a frequency comb, \textcolor{black}{while the} mixing between \textcolor{black}{the} neurons is realized by electro-optic phase modulators \cite{Butschek2022}. This architecture allows for a relatively easy\textcolor{black}{-}to\textcolor{black}{-}realize optical output layer, as \textcolor{black}{the output} weights can be applied \textcolor{black}{to the} comb lines \textcolor{black}{using} a programmable spectral filter, \textcolor{black}{while} the nonlinear summation of the weighted neurons can be executed by a photodiode\textcolor{black}{. The photodiode} measures the total intensity of the weighted frequency comb and introduces a quadratic nonlinearity. This technique, already employed in \cite{Butschek2022} to generate the output signals with optical weighting, allows us to use the output of a reservoir as \textcolor{black}{an} input to a second one without leaving the analog domain. \textcolor{black}{Here,} we also fully exploit the frequency degree of freedom of light by using the same hardware for implementing multiple reservoirs simultaneously, each one working in a different frequency band. In particular,  we report two simultaneous RC computations and demonstrate that combining \textcolor{black}{these computations} in a deep fashion improves \textcolor{black}{the overall} performance \textcolor{black}{as} compared to using \textcolor{black}{two independently running parallel} reservoirs. We test two strategies for optimizing the interconnections between \textcolor{black}{the} layers in the deep configuration. In the first \textcolor{black}{(}simpler\textcolor{black}{)} approach\textcolor{black}{,} we only adjust the strength of the connection\textcolor{black}{s, whereas} in the second approach\textcolor{black}{,} we optimize the connections using the Covariance Matrix Adaptation Evolution Strategy (CMA-ES)\cite{hansen2006cma}.
 \textcolor{black}{To our surprise}, we find that both approaches yield comparable results.

In Sec.\ \ref{sec:methods} we present the algorithms, the experimental setup and the benchmarking methods; in Sec.\ \ref{sec:results} we present and discuss results; finally, in Sec.\ \ref{sec:conclusion} we present conclusions and outlooks for this work.

\section{Methods}\label{sec:methods}
\subsection{Algorithms}


\begin{figure}
  \centering
  \includegraphics[width=\textwidth]{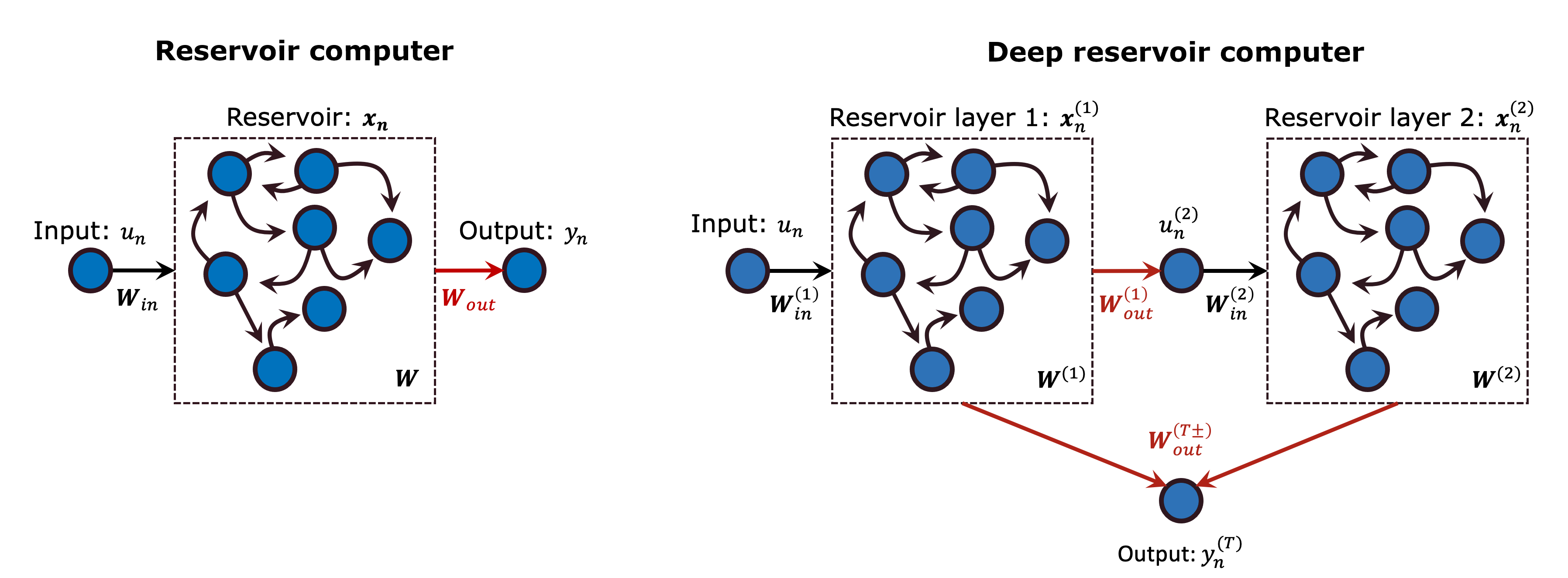}
  \caption{Left panel: standard reservoir computing scheme. Right panel: deep reservoir computing scheme. The weights in black are fixed, while the weights in red are trained.}\label{fig:configurations}
\end{figure}

\subsubsection{Reservoir computing}\label{sec:RC}
A reservoir computer (RC, see left panel of Fig.\ref{fig:configurations}) \cite{Jaeger04} is a recurrent neural network composed of three layers: the input layer, the reservoir layer, and the output layer. Only the output weights are trained, while the input and internal weights are fixed and \textcolor{black}{remain untrained}. 

The experimental system is based on the frequency multiplexing RC scheme described in \cite{Butschek2022}. \textcolor{black}{The n}eurons are encoded in the complex amplitudes of the lines of a frequency comb and \textcolor{black}{the} neuronal interconnections are realized via frequency-domain interference \textcolor{black}{that provides a power exchange between the lines}. The electric field in the reservoir can\textcolor{black}{,} thus\textcolor{black}{,} be expressed as 
\begin{equation}
E(t) = \sum_k x_{k}(t) \exp \left( i(\omega + k\Omega)t \right),\label{Eq:E(t)}
\end{equation}
where $\omega$ is the center frequency of the comb, $\Omega$ the frequency spacing between the comb lines, and  $x_{k}(t) $ are the slowly varying amplitudes of the comb lines \textcolor{black}{that} encode neuron information. To describe the RC application more conveniently, we focus on the $N$ most central lines of the comb, which are the ones encoding information. Moreover, we group the amplitude of these lines in a $N$ dimensional complex vector $\Xrc{n}$ that evolves in slow, discrete time $n$. The discrete timescale corresponds to the discrete evolution of the RC states.

The RC based on frequency multiplexing uses nonlinear input and output layers and a linear reservoir (which is a powerful architecture, as demonstrated in \cite{vinckier2015high}). It can be described by the  evolution equations:
\begin{align}
    \Xrc{n} &= \Wrc{}\cdot\Xrc{n-1} + \Wrc{in}\cdot\frc{in}\left(\usigrc{n}\right), \label{eq:ONE}\\
    \Yrc{n} &= \Wrc{out}^T \cdot  \vert \Xrc{n}\vert^2,\label{eq:TWO}
\end{align}
where
$\usigrc{n}$ (a real scalar) is the input signal to the reservoir at timestep ${n}$, $\Yrc{n}$ (a real scalar) is the output signal of the reservoir at timestep $n$, $\Wrc{}$ is a complex $N\times N$ matrix representing the internal connections of the reservoir,
$\Wrc{in}$ is a complex $N$ dimensional vector representing the input-to-reservoir connections, 
\textcolor{black}{
$\Wrc{out}$ is a vector of $N$ real readout weights with $\ ^T$ denoting the transpose, and $\vert \Xrc{n}\vert^2$ is the vector obtained by taking the norm square of $\Xrc{n}$ elementwise.
The output weights are optimized using ridge regression so that the output $y_n$ approximates the desired output as well as possible.}

In our implementation, the  input signal is \textcolor{black}{provided} through a Mach-Zehnder modulator operating in the negative quadrature point. \textcolor{black}{Hence,} the input nonlinearity $\frc{in}$ is given by the modulator transfer function:
\begin{equation}\label{eq:f_in}
    \frc{in}(u) = E_0\cdot\sin\left(\gamma\cdot u\right),
\end{equation}
where $E_0$ represents the input radiation amplitude and $\gamma$ is the driving strength of the electrical signal to the modulator. 

\textcolor{black}{
Eq.\ \eqref{eq:TWO}  can be implemented  by measuring each component of 
$\vert \Xrc{n}\vert^2$ and then carrying out the scalar product offline, i.e.\ on a digital computer. This is the method used in the present work. 
However, we note that the output  $\Yrc{n}$ can  also be obtained directly in the analog domain using the following procedure \cite{Lupo2021,Butschek2022}. The optical signal is sent to a programmable spectral filter with two outputs yielding the two signals $\Wrc{out}^{+}\cdot\Xrc{n}$ and $\Wrc{out}^{-}\cdot\Xrc{n}$ (each given by a
complex $N$-dimensional vector representing the the comb line amplitudes)
where 
 $\Wrc{out}^+$ and $\Wrc{out}^-$ are $N\times N$ diagonal matrices with positive real coefficients  corresponding respectively to the square root of the positive and negative elements of $\Wrc{out}$. These two signals are then sent to two photodiodes that measure their total power and the difference of the powers is computed. Accordingly, the output reads as:
\begin{equation}
  \Yrc{n} = 
   \left \vert \Wrc{out}^{+}\cdot\Xrc{n}\right\vert^2
    -\left \vert \Wrc{out}^{-}\cdot\Xrc{n}\right\vert^2 ,\label{eq:TWOB}
\end{equation}
  where $\vert \cdot\vert^2$ denotes taking the norm square of a vector. 
  }

\subsubsection{Deep Reservoir Computing}\label{sec:deepRC}
A deep reservoir computer (deep-RC, see right panel of Fig.\ \ref{fig:configurations}) is a \textcolor{black}{stack} of RC layers connected in series. The deep-RC output signal is a linear combination of neuron values of each reservoir. The hierarchy introduced by the serial connection enhances the network performance \textcolor{black}{as} the different reservoirs can have \textcolor{black}{different} dynamics, thus enriching the states of the full deep-RC.

A deep-RC composed of $N_\text{layers}$ layers, each one comprising $N$ neurons, as implemented in our system, is described by the set of equations:
\begin{align}
    \X{n}{1} &= \W{}{1}\cdot\X{{n}-1}{1} + \W{in}{1}\cdot\frc{in} \left(\usigrc{n}\right), & \label{eq:deepRC0}\\
    \X{n}{i} &= \W{}{i}\cdot\X{{n}-1}{i} + \W{in}{i}\cdot\frc{in} \left(\usigrc{n}^{(i)} \right),\quad&i&=2,\ldots, N_\text{layers} \label{eq:deepRC1}\\
  \usigrc{n}^{(i+1)} &= \left\vert\W{out}{i}\cdot\X{n}{i}\right\vert^2,\quad&i&=1,\ldots, N_\text{layers}-1 \label{eq:deepRC2}\\    
    \Y{n}{\text{A}} &= 
 \Wrc{out}^{\text{(A)}\ T} \cdot  \vert \Xrc{n}^{\text{(A)}}\vert^2    
   \label{eq:deepRCout}
\end{align}
where the superscript $(i)$, $1\leq i\leq N_\text{layers}$, identifies the reservoir layer. \textcolor{black}{As} before, 
$\W{}{i}$ is a complex $N\times N$ matrix representing the internal connections of the $i$-th reservoir layer, 
$\W{in}{i}$ is a complex $N$ dimensional vectors representing the input connections of the $i$-th reservoir layer,
$\W{out}{i}$ is a $N\times N$ diagonal \textcolor{black}{matrix} with positive real coefficients representing the output connections of the $i$-th layer
\textcolor{black}{and $\vert \cdot  \vert$ in Eq. \eqref{eq:deepRC2} denotes the norm square of the argument, which is a vector.
In the experiment, we use a two-layer configuration, i.e.\ $N_\text{layers}=2$,} but the equations \textcolor{black}{easily generalize} to more layers. The first reservoir layer is driven by the input time series $u_n$, while the \textcolor{black}{subsequent} reservoir layers are driven by the output\textcolor{black}{s} $\usigrc{n}^{(i)}$ of the \textcolor{black}{preceding layers (}Eq.\ \eqref{eq:deepRC2}\textcolor{black}{)}. Note that in our implementation\textcolor{black}{,} the connections \textcolor{black}{between the} consecutive layers only consists of positive weights, contained in the diagonal of $\W{out}{i}$. \textcolor{black}{This} is why there is only a single term on the right hand side of Eq.\ \eqref{eq:deepRC2} (as compared to Eq.\ \eqref{eq:TWOB}). \textcolor{black}{We note that these equations do not account for possible delays between the consecutive RC layers introduced by the experimental setup (e.g., the length of optical or electrical connections). This corresponds to the situation in our experiment, in which these delays are compensated in the digital postprocessing. We note that such delays can in principle always be compensated for by adding optical or electrical delay lines of appropriate length.}

The deep-RC output, $\Y{n}{\text{A}}$, is obtained by combining the states from \textcolor{black}{all} layers, i.e.\ by 
\textcolor{black}{ taking a linear combination of the intensities of all the comb lines with a photodiode.
To express this, we have defined $\X{n}{\text{A}}=\left(\X{n}{1},  \X{n}{2},\ldots,  
\mathbf{x}_{n}^{(N_{\text{layers}})}
\right)$ as the complex vector of size $N_\text{layers}\cdot N$ representing the full deep-RC state at timestep $n$,
and $\W{out}{\text{A}}$ as the vector of 
 $ (N_\text{layers}\cdot N)$ real output weights. The output weights are optimized using ridge regression.}

Note that the interconnection between \textcolor{black}{the consecutive layers} (say layers $i$ and $i+1$) is determined by $3N$ real parameters: the $N$ positive real elements of the diagonal matrix $\W{out}{i}$, and both the real and imaginary parts of the $N$ elements of the vector $\W{in}{i+1}$. Of these, only $N$  elements of the diagonal matrix $\W{out}{i}$ can be tuned in our experimental setup. (This is to be compared with the proposal of \cite{Gallicchio2017}, \textcolor{black}{where} the interconnection is given by a $N\times N$ random matrix, whose spectral radius is tuned. \textcolor{black}{ More advanced algorithms and topologies, such as presented in \cite{freiberger2019improving,Freiberger2019Training,Nakajima2022}, aim to achieve more freedom in tuning interconnections.)}

\subsection{Experimental setup}

\begin{figure}
    \centering
    \includegraphics[width=0.8\textwidth]{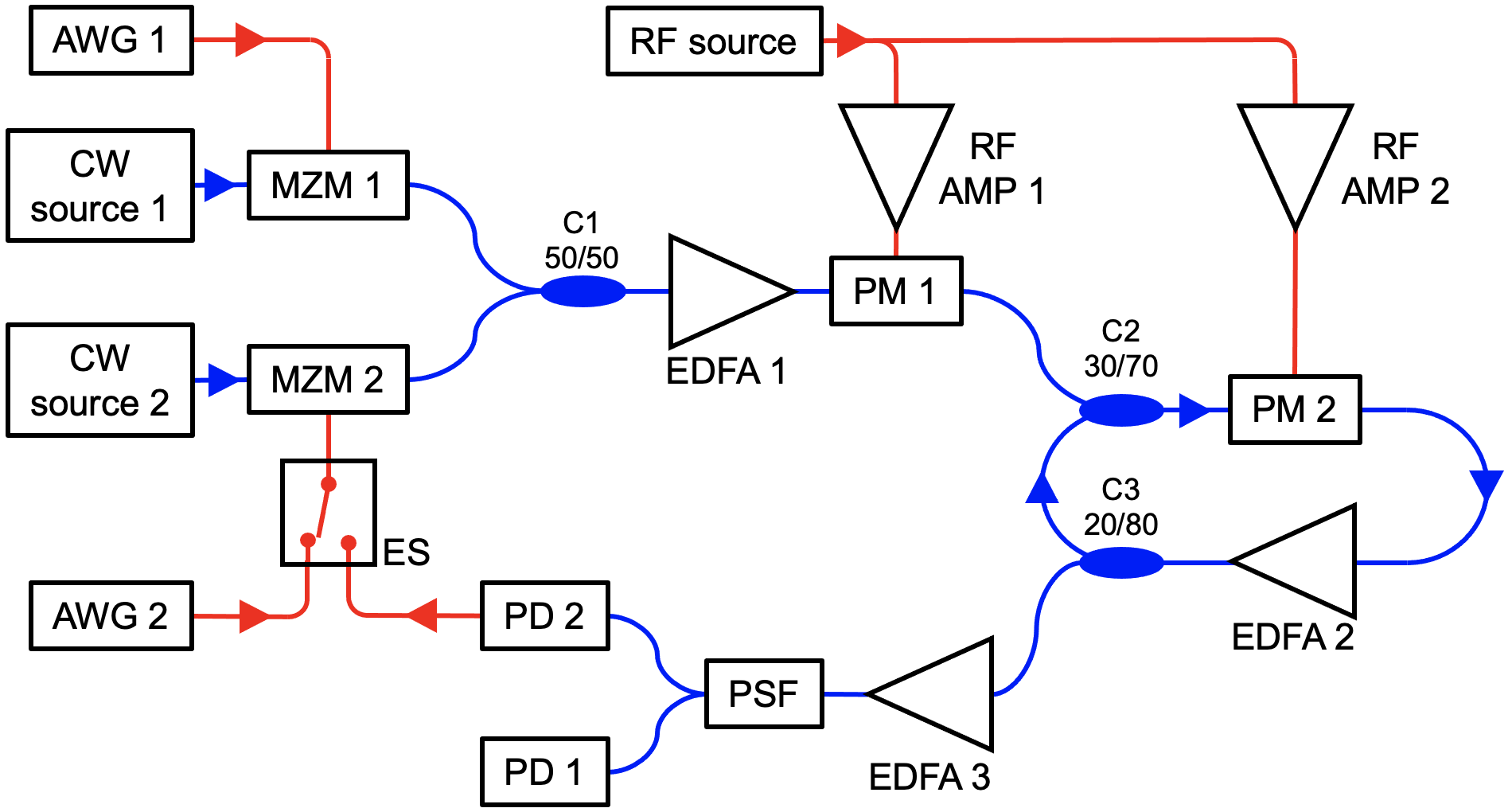}
    \caption{Experimental setup. Optical connections are in blue, electrical connections in red.
MZM: Lithium Niobate Mach-Zehnder modulator;
AWG: arbitrary waveform generator;
C: fiber couplers;
EDFA:  Erbium-doped-fiber amplifier;
PM: phase modulator; 
RF source: radio frequency source at frequency $\Omega$;
RF AMP: radio frequency amplifiers;
PSF: programmable spectral filter;
PD: photodiode;
ES: electric switch.}
    \label{fig:setup}
\end{figure}

\begin{figure}
    \centering
    \includegraphics[width=1\textwidth]{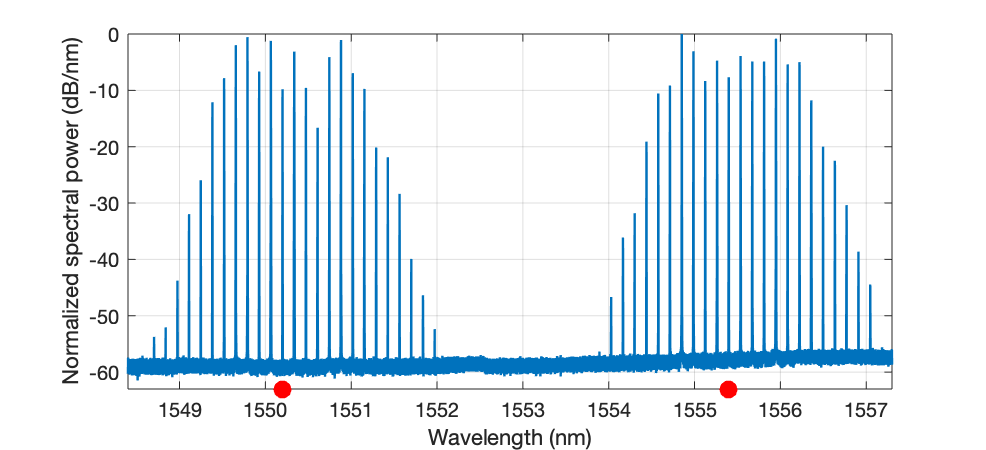}
    \caption{Normalized spectral power of the radiation as measured at the output of the fiber loop, after coupler C 3. Red markers indicate the input wavelengths $\lambda_1=1550.2$ nm and $\lambda_2=1555.4$ nm.}
    
    \label{fig:comb}
\end{figure}

\paragraph{Experimental Setup.}

The experimental system  is based on \cite{Butschek2022}, modified \textcolor{black}{such that it supports} two RC computations at the same time. 
 
Fig.\ \ref{fig:setup} \textcolor{black}{shows} the schematic of the experiment. All fiber connections and couplers are single-mode and polarization-maintaining. We employ two continuous-wave laser sources (CW source 1 and CW source 2) at wavelengths $\lambda_1=1550.2$ nm and $\lambda_2=1555.4$ nm. The laser outputs are modulated by two Mach-Zehnder modulators (MZM 1 and MZM 2). Both MZMs are biased to operate in the negative quadrature point (bias controllers are not shown in Fig.\ \ref{fig:setup}). The transfer functions of MZM 1 and MZM 2 define the input nonlinearities of the two RC layers, $\frc{in}$ in Eqs.\ (\ref{eq:f_in}, \ref{eq:deepRC0}, \ref{eq:deepRC1}). MZM 1 is driven by an arbitrary waveform generator (AWG 1) which supplies the input signal $\usig{n}{1}.$ MZM 2 can be driven by a second arbitrary waveform generator (AWG 2) or the output of \textcolor{black}{another} photodiode (PD 2). The modulated signals are merged together in a 50/50 fiber coupler (C 1) and then injected into an Erbium-doped-fiber amplifier (EDFA 1). EDFA 1 raises the total power to $9$ dBm \textcolor{black}{that is} equally distributed between \textcolor{black}{the} signals. After the amplification, \textcolor{black}{both} signals are injected \textcolor{black}{into} a phase modulator (PM 1). PM 1 is driven by a sinusoidal radio-frequency signal (frequency $\Omega\approx17$ GHz, power P1$\approx30$ dBm). The radio-frequency signal is generated by an RF clock (RF source) and amplified by an RF amplifier (RF AMP 1). The phase modulation provided by PM 1 generates two frequency combs centered \textcolor{black}{at} $\lambda_1$ and $\lambda_2$ (Fig.\ \ref{fig:comb}). The spacing of the comb lines is equal to $\Omega$ and the number of lines depends on P1. In our implementation, PM 1 provides approximately $20$ usable comb lines per comb, i.e., $20$ neurons. \textcolor{black}{Both frequency} combs constitute the input stimuli for the two reservoir networks. The amplitude of each line determines how strongly the input signal is coupled to the particular neuron encoded in that line. Hence, the distribution of (complex) amplitudes among the comb lines defines two vectors of \textcolor{black}{the} input-to-reservoir weights, $\W{in}{1}$ and $\W{in}{2}$. \textcolor{black}{Both} frequency combs are injected in a fiber loop through a 30/70 coupler (C 2). The fiber loop is 15 meters long, corresponding to a roundtrip frequency of approximately $20$ MHz. The input signals are synchronized with the roundtrip time of the loop \textcolor{black}{such that} each timestep of the input signals entirely fills the loop. Hence, the processing frequency of our system is fixed by the cavity length and is approximately $20$ MHz. The fiber loop contains a second phase modulator (PM 2) and an optical amplifier (EDFA 2). PM 2 is driven by a signal 
generated by the same RF source \textcolor{black}{as PM 1, but it undergoes a different amplification (RF AMP 2). Hence,} it has the same frequency but a different power, P2$\approx$20 dBm\textcolor{black}{.} The phase modulation provided by PM 2 creates frequency interference among the lines of the same comb, thus implementing the (complex-weighted) \textcolor{black}{connection between} the neurons of the same reservoir. EDFA 2 compensates for the losses in the loop. The transformation of the combs over a roundtrip, including the effects of phase modulation, amplification, and dispersion (which acts differently on each comb line/neuron) define the matrices $\W{}{1}$ and $\W{}{2}$. The amplitudes of \textcolor{black}{both} combs at each roundtrip ${n}$ provide  the states of the two reservoirs $\X{n}{1}$ and $\X{n}{2}$. 

\textcolor{black}{A part} of the circulating radiation is extracted by a 20/80 fiber coupler (C 3), amplified by EDFA 3, and directed to the readout circuit. The readout  consists of a multi-channel programmable spectral filter (PSF, \textit{Coherent II-VI Waveshaper}, \textcolor{black}{with resolution $0.01$ dB}) and two photodiodes (PD 1 and PD 2), measuring each of the two PSF outputs. 
The first PSF channel, connected to PD 1, is employed to measure the evolution of both reservoirs. The measurement procedure consists of selecting a single comb line \textcolor{black}{per time, by} setting a band-pass filter \textcolor{black}{on the PSF channel,} and \textcolor{black}{recording the intensity of this comb line by PD 1}. At the end of the procedure, the intensities of all comb lines, i.e.\ the norm square of the components of vectors $\X{n}{1}$ and $\X{n}{2}$, are recorded on a computer. \textcolor{black}{Ridge regression is employed to train the output weights 
$ \W{out}{\text{A}} $ (with a regularization parameter of $10^{-5}$, considering neuron signals of the order of 1).} 
The output of the reservoir is then obtained by multiplying \textcolor{black}{measured line} intensities by \textcolor{black}{trained} output weights. Note that \textcolor{black}{the training can only be realized with the support of a digital computer, while the application of the output weights can} be realized \textcolor{black}{optically in the analog domain (Eq. \eqref{eq:TWOB}).}

\paragraph{Operation modes.}

We use two operation modes: ``deep" and ``independent". 

In \textcolor{black}{the} deep-RC mode, the second channel of the programmable spectral filter is configured to select and transmit only the comb centered \textcolor{black}{at} $\lambda_1$ after having applied an attenuation mask $\W{out}{1}$. Consequently, PD 2 measures the signal $\usigrc{n}^{(2)}=\left\vert\W{out}{1}\cdot\X{n}{1}\right\vert^2$. The output of PD 2 drives MZM 2, and thus constitutes the input of the second RC  \textcolor{black}{at $\lambda_2$}. In this configuration, the system is a two\textcolor{black}{-}layer deep-RC, as described in subsection \ref{sec:deepRC}.

In \textcolor{black}{the} independent mode, \textcolor{black}{both} RC computations are decoupled by driving MZM 2 through a second, independent, arbitrary waveform generator AWG 2 (the second channel of PSF and PD 2 are deactivated). \textcolor{black}{Thus, the} computations do not interact with each other and are carried out independently. 

The selection of the computation mode, deep or independent, is made by flipping an electric switch that selects whether MZM 2 is driven by PD 2 or by AWG 2, as illustrated in Fig.\ \ref{fig:setup}.

\paragraph{Stabilization.}
The  experimental setup is sensitive to acoustic noise and thermal drift.
To limit these effects, the optical loop, including PM 2 and EDFA 2, is mounted inside an insulated box on an optical table.
Furthermore, two PID controllers piezo-tune the emission wavelengths of \textcolor{black}{both} laser sources in order to fix the operating condition to a certain point in the loop transfer function. The PID controllers are fed by the intensity of the reflection of each comb at the entrance of the loop, \textcolor{black}{i.e.\ at} the coupler C 2. This requires two auxiliary photodiodes and spectral filters (not represented in Fig.\ \ref{fig:setup}).

\subsection{Benchmark tasks}
We selected two benchmark tasks, the first consisting of the prediction of the evolution of a chaotic time series, and the second one consisting of the compensation of the distortion produced in a nonlinear communication channel.

The time series prediction task is based on the infrared laser dataset of the Santa Fe Time Series Competition \cite{Weiss95}. The time series $u_t$ is supplied as input, and the task consists of producing $u_{t-\tau}$, with $-5\leq\tau\leq+5$. Note that when the timeshift $\tau$ is negative, the task consists of remembering the past, while when the timeshift $\tau$ is positive, the task consists of predicting the future. The accuracy is expressed in terms of \textcolor{black}{the} normalized mean square error (NMSE) between the target signal and the produced output. When running this benchmark, the training set is composed of 6000 timesteps, and the testing set is composed of 2500 timesteps (this is a standard 70\%-30\% repartition). We discard the first 500 timesteps of \textcolor{black}{the} reservoir output to avoid operating in a transient \textcolor{black}{phase}. 

The nonlinear channel compensation task was first used in the RC community in \cite{Jaeger04}. A random signal composed of four different symbols is propagated along a simulated channel exhibiting nonlinearity, noise, and memory about past inputs. The task consists in reconstructing the original input given the channel output. \textcolor{black}{The} performance is evaluated for different signal-to-noise ratios (SNR) in the range \textcolor{black}{of} $[8\text{dB},\text{ }32\text{dB}]$. The results are expressed in terms of \textcolor{black}{the} Symbol Error Rate (SER), i.e.,\ the ratio of wrongly reconstructed output symbols over the total number of transmitted symbols. When running this benchmark, the training set is composed of 14000 timesteps, and the testing set is composed of 30000 timesteps. We discard the first 1000 timesteps of \textcolor{black}{the} reservoir output to avoid operating in a transient \textcolor{black}{phase}. Note that, \textcolor{black}{contrary to the time-series benchmark relying on a limited dataset}, \textcolor{black}{the nonlinear channel} dataset can be easily generated \textcolor{black}{on the fly}. This is why we employed a larger amount of data points for the initial wash-out and the testing.

Every benchmark result has been validated through a 100-steps cross-validation, meaning that the points belonging to the train and test datasets have been selected at random for 100 times and \textcolor{black}{the} results have been averaged.

\subsection{Tested configurations}\label{sec:tested}

\begin{figure}
  \centering
  \begin{subfigure}[b]{0.3\textwidth}
      \caption{}
    \label{fig:shallow_scheme}
    \includegraphics[width=\linewidth]{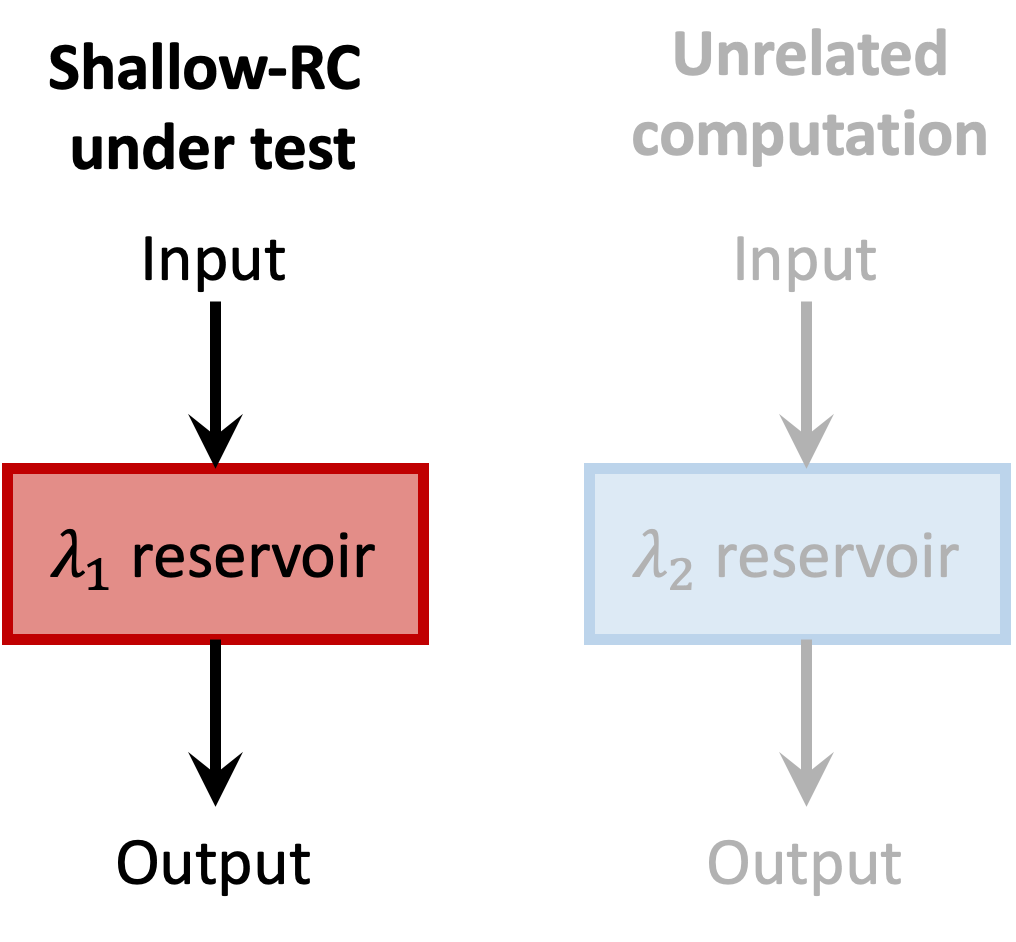}
  \end{subfigure}
  \hfill
\begin{subfigure}[b]{0.3\textwidth}
    \caption{}
    \label{fig:parallel_scheme}
    \includegraphics[width=\linewidth]{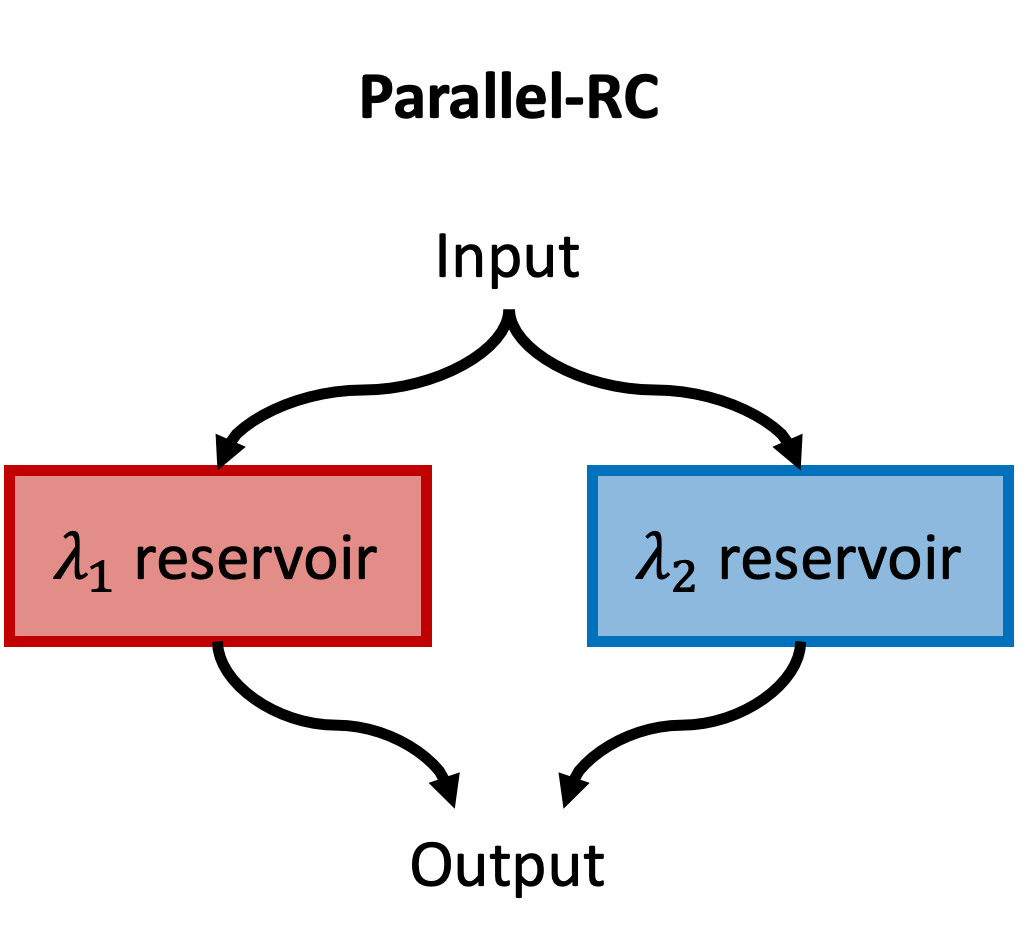}
  \end{subfigure}
  \hfill
\begin{subfigure}[b]{0.3\textwidth}
    \caption{}
    \label{fig:deep_scheme}
    \includegraphics[width=\linewidth]{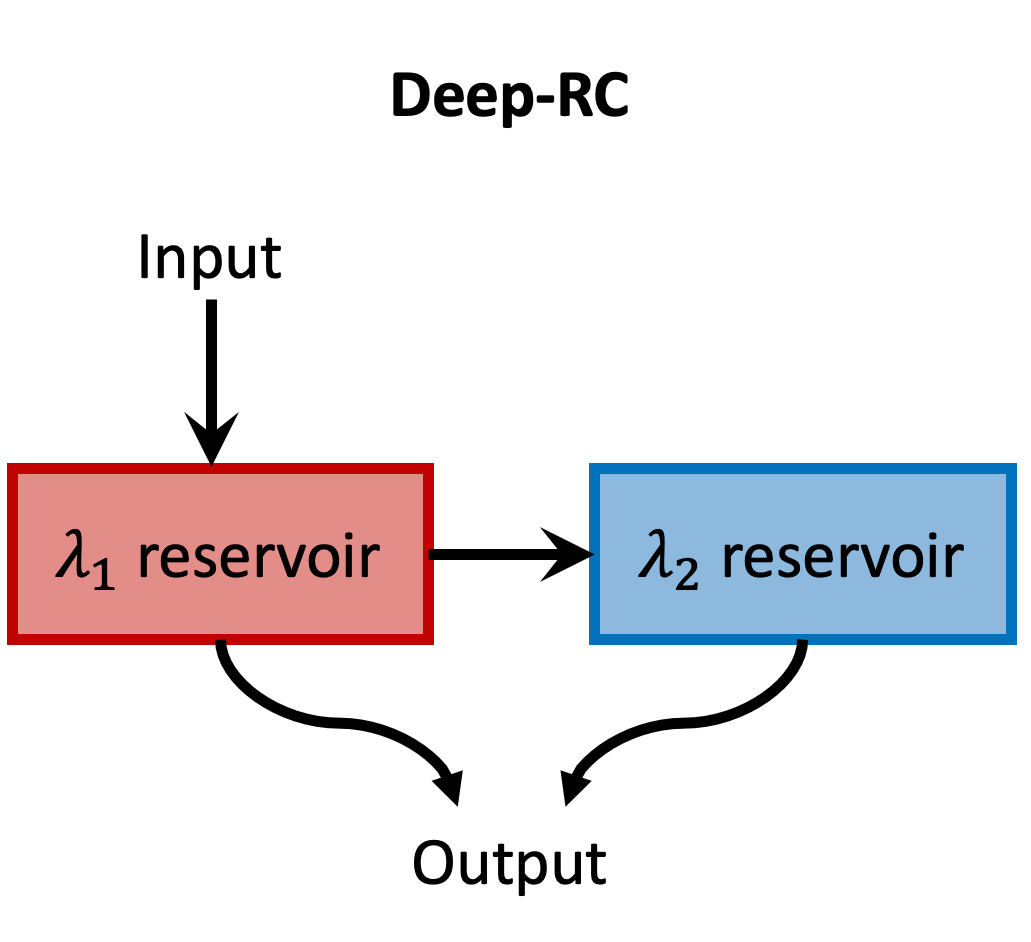}

  \end{subfigure}
  
  \caption{The three tested configurations for the two independent RCs. 
  `` Reservoir $\lambda_1$" is encoded in the frequency comb centered around $\lambda_1$, while ``reservoir $\lambda_2$" is encoded in the frequency comb centered around $\lambda_2$. Both reservoirs are executed on the same photonic substrate.  (a) Shallow-RC: one of the two reservoirs performs the benchmark task as a traditional RC, while the other reservoir processes a different time series in parallel. (b) Parallel-RC: both reservoirs process the same input time series, but their dynamics are decoupled from each other. A single output layer is trained, which combines signals from both reservoirs. (c) Deep-RC: two reservoirs constitute two layers of a deep-RC.}\label{fig:schemes}
\end{figure}

Our photonic system supports two RCs that operate simultaneously, either independently\textcolor{black}{, i.e.\ in parallel,} or connected in series. We  evaluated the performance of three different configurations, described in Fig.\ \ref{fig:configurations}. First, \textcolor{black}{we consider} a ``shallow-RC" configuration (Fig.\ \ref{fig:shallow_scheme}) \textcolor{black}{where} only one of two independent RCs executes the benchmark task, constituting a ``traditional"  RC as described in Sec.\ \ref{sec:RC}. In this configuration, the second RC processes a different, not evaluated, computation with the purpose of simulating a parallel\textcolor{black}{-}computation scenario where two different tasks are performed at the same time. Second, \textcolor{black}{we study} a ``parallel-RC" configuration (Fig.\ \ref{fig:parallel_scheme}) where both independent RCs execute the same task in an uncorrelated way, and a single output layer is connected to both reservoirs. This constitutes a ``non-deep" way of using the full computational capabilities of the system on a single task. Third, \textcolor{black}{we consider} a ``deep-RC" configuration (Fig.\ \ref{fig:deep_scheme}) where two independent RCs are connected in series as described in Sec.\ \ref{sec:deepRC}. 

In addition, in the deep-RC configuration, we used two different methods to tune the  weights $\W{out}{1}$, i.e.\ the attenuations applied by the PSF, that  determine the connection from the first RC layer to the second one. In the first, simplest approach, we apply the same attenuation to all comb lines, corresponding to $\W{out}{1} = \textrm{diag}(\alpha)$, and we optimize the overall attenuation $\alpha^2$ by sweeping it in the range \textcolor{black}{of} $[-20\text{dB},\text{ }0\text{dB}]$.
In the second approach, we optimize all the coefficients of $\W{out}{1}$ by using the Covariance Matrix Adaptation Evolution Strategy (CMA-ES) optimization algorithm \cite{hansen2006cma}. \textcolor{black}{CMA-ES is a standard tool for continuous black-box optimization, already used in the context of reservoir computing in \cite{Freiberger2019Training}. The algorithm consists in sampling possible solutions from a multivariate Gaussian distribution whose parameters (mean and covariance) are tuned based on the performance of the solutions sampled in the previous epochs. In our case, \textcolor{black}{the} optimization runs over six epochs, \textcolor{black}{using} a population of 13 sampled solutions per epoch. 
For each choice of weights $\W{out}{1}$, we optimize the output weights and then use the performance on the corresponding RC as a measure of fitness.
Independent of the strategy (sweeping of $\alpha^2$ or CMA-ES), the purpose of the optimization is to find the configuration \textcolor{black}{that} maximizes the network performance.}

\begin{figure}
     \centering
    \includegraphics[width=\textwidth]{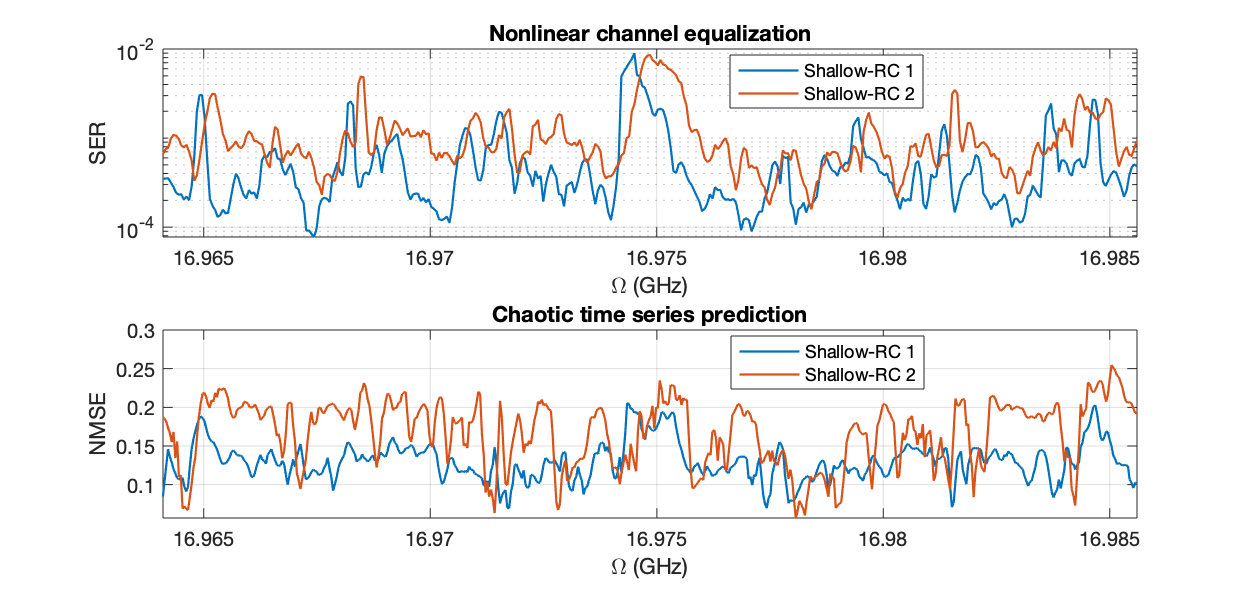}
        \caption{Performance of the reservoir computers in shallow-RC configuration as a function of $\Omega$ on the channel equalization task (top) and the Santa-Fe time series prediction task for prediction 1 timestep ahead (bottom) . The complex dependence on $\Omega$ is due to the dispersion in the optical fiber. The dispersion is also the reason why the dependence on $\Omega$ is different for RC-1 and RC-2, as they use frequency combs centered on different wavelengths. (As these plots are time-consuming to obtain, a reduced number of comb lines $N=14$ was used).        
        }
        \label{fig:resultsB}
\end{figure}

Finally, to improve the reservoir computing performance, we tune the comb line spacing $\Omega$ to the best-performing value for each task. The fiber loop constitutes a spectral interferometer and exhibits, due to dispersion in the fiber, a complex behavior strongly dependent on $\Omega$. This is illustrated in Fig.\ \ref{fig:resultsB}, where the performance \textcolor{black}{of the} shallow-RC configuration for two different tasks is plotted as a function of $\Omega$ for both reservoirs. 

\section{Results}\label{sec:results}

\begin{figure}
     \centering
     \begin{subfigure}[b]{0.49\textwidth}
         \centering
         \includegraphics[width=\textwidth]{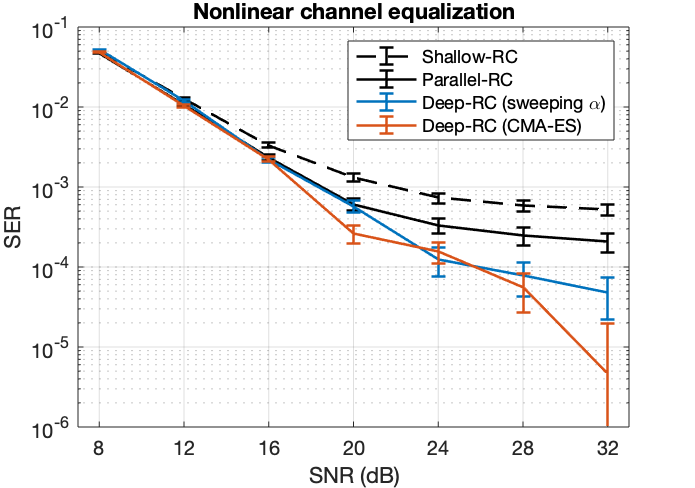}
         \caption{}
         \label{fig:nlc_res}
     \end{subfigure}
     \hfill
     \begin{subfigure}[b]{0.49\textwidth}
         \centering
         \includegraphics[width=\textwidth]{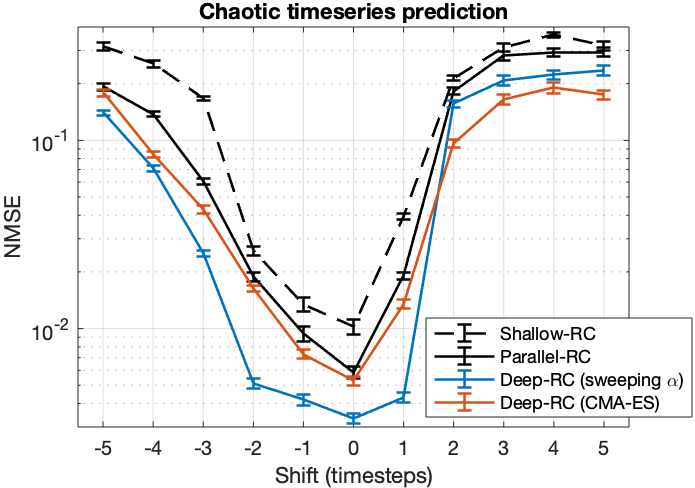}
         \caption{}
         \label{fig:santafe_res}
     \end{subfigure}
        \caption{Experimental results for the three operation modes (shallow-RC, parallel-RC, and deep-RC) on the two selected benchmark tasks: nonlinear channel equalization (a) and chaotic time series prediction (b). Deep-RC results are shown for both optimization methods presented in the text (uniform optimized attenuation $\alpha$ and CMA-ES). Error-bars represent the score standard deviation measured in cross-validation phase. Results in (a) are expressed as symbol error rate (SER) vs. signal-to-noise ratio (SNR). Results in (b) are expressed as normalized mean square error (NMSE) vs. shift of the target time series with respect to the input one. When the shift is positive, the task consists of predicting the future; when the shift is zero, the task consists of reproducing the present input; when the shift is negative, the task consists of reproducing the past.}
        \label{fig:results}
\end{figure}



\textcolor{black}{The results} of the benchmark tasks are \textcolor{black}{shown} in Fig.\ \ref{fig:results} for the three operation modes: shallow-RC, parallel-RC, and deep-RC. In this figure, the deep-RC results are shown for both optimization techniques described in Sec.\ \ref{sec:tested}. \textcolor{black}{The nonlinear-channel equalization} results (Fig.\ \ref{fig:nlc_res}) show the expected decrease of \textcolor{black}{the} symbol error rate (SER) \textcolor{black}{with increasing} signal-to-noise ratio (SNR). \textcolor{black}{This is because additional noise raises the complexity of the task, which eventually makes correcting the signal distortion impossible.} For high-SNR values, both shallow-RC and parallel-RC SER scores saturate, while \textcolor{black}{the} deep-RC SER score maintains an exponential decay for increasing SNR values. For every SNR value, \textcolor{black}{the} deep-RC always \textcolor{black}{performs} better, followed by the parallel-RC and finally shallow-RC. A similar behavior is found in the results of the chaotic time series prediction task (Fig.\ \ref{fig:santafe_res}). 

Two trends are clearly visible \textcolor{black}{in} Fig. \ref{fig:results}.
First, \textcolor{black}{the} parallel-RC systematically outperforms \textcolor{black}{the} shallow-RC. Indeed\textcolor{black}{,} since two parallel RCs perform different computations (as is evident from  Fig. \ref{fig:resultsB}), using both reservoirs in parallel should perform at least as well as using a single reservoir. 
Second, \textcolor{black}{the} deep-RC outperforms \textcolor{black}{the} parallel-RC in every test we conducted. Both configurations exploit the same number of neurons and \textcolor{black}{differ only by their topology.} \textcolor{black}{Thus,} we conclude that the serial configuration in \textcolor{black}{a} deep-RC really boosts \textcolor{black}{RCs overall} performance.

We observe that \textcolor{black}{both} optimization techniques for the inter-layer connection perform comparably, with the simpler algorithm sometimes outperforming the CMA-ES algorithm. We identified two reasons for this behavior. First, the CMA-ES algorithm could get stuck \textcolor{black}{in a local minima}. Second, the search for the optimal set of weights could be affected by slow drifts in the operating conditions of the deep-RC. \textcolor{black}{For example, as reported in \cite{Butschek2022}, the transfer function of the fiber loop constituting the reservoir is strongly sensitive to temperature changes. \textcolor{black}{This is because the thermally induced fiber-lenght variation affects the relative phases of the comb lines, thus, changing the dynamics of the system}. The complexity of the system is well captured \textcolor{black}{in} Fig.\ \ref{fig:resultsB} which shows how a small change in a parameter significantly affects the performance. Improved stabilization should resolve this issue in the future.}

\section{Conclusion}\label{sec:conclusion}

We presented a fully analog photonic implementation of a \textcolor{black}{two-layer} deep reservoir computer. The connection between \textcolor{black}{the} layers is performed in the analog domain, with no processing or storing on a digital computer. The presented implementation also allows for two independent RC computations to be executed at the same time. We found that the deep-RC configuration, obtained by connecting \textcolor{black}{two RCs in series,} performs better than a parallel-RC configuration, where two RCs \textcolor{black}{process the same input data without interacting.}

The reported experiment has only two layers, but deeper schemes are in principle possible. New layers can be added to the deep-RC by using more than two lasers, provided that the generated combs do not overlap each other. The C band could host $10$ parallel computations (considering combs \textcolor{black}{with widths of $3$ nm}, see Fig.\ \ref{fig:comb}). These $10$ parallel computations could be employed to constitute a single $10$-layers deep-RC, or even multiple deep-RC running in parallel, each one composed of fewer layers. On the other hand, broader combs would be able to encode more neurons in each reservoir. Thus, a balance between the number of layers and the number of neurons per layer has to be searched for. In any case, integrating (partially or entirely) the experiment, as proposed in \cite{Kassa2018}, could be a route to scaling up the system while simplifying its stabilization.

\textcolor{black}{We note that, for simplicity, in the present experiment the final output $   \Yrc{n}$ was obtained digitally by carrying out the linear combination as described in Eq. \eqref{eq:TWO}. However, as described in Eq. \eqref{eq:TWOB} and reported in our previous works \cite{Lupo2021,Butschek2022}, the output can also be obtained in the analog domain without loss of performance.}

\textcolor{black}{
Although we explored two strategies for optimizing the interconnection between the two deep-RC layers, many ideas are still to be tested (see e.g., \cite{Gallicchio2017,wright2022deep,Nakajima2022}) and could be the object of further investigation. }

In summary, developing deep architectures for neuromorphic photonic computing is a highly promising avenue for increasing both the complexity of the tasks that can be solved and the system performance. However, the presence of analog-to-digital or digital-to-analog converters strongly affects \textcolor{black}{the} power consumption and footprint, hence it is to be avoided. We have demonstrated that this is possible for photonic deep reservoir computing.

\section*{Funding}
The authors  acknowledge financial support from the European Union’s Horizon 2020 research and innovation program under the Marie Skłodowska-Curie grant agreement  860830 (POST DIGITAL), and from the FWO and F.R.S.-FNRS Excellence of Science (EOS) programme grant 40007536. 

\section*{Disclosures}
The authors declare no conflicts of interest.

\section*{Data availability}
Data underlying the results presented in this paper are not publicly available at this time but may be obtained from the authors upon reasonable request.

\printbibliography

@article{xu202111,
	author = {Xu, Xingyuan and Tan, Mengxi and Corcoran, Bill and Wu, Jiayang and Boes, Andreas and Nguyen, Thach G and Chu, Sai T and Little, Brent E and Hicks, Damien G and Morandotti, Roberto and others},
	journal = {Nature},
	number = {7840},
	pages = {44--51},
	publisher = {Nature Publishing Group UK London},
	title = {11 TOPS photonic convolutional accelerator for optical neural networks},
	volume = {589},
	year = {2021}}

@article{silver2018general,
	author = {Silver, David and Hubert, Thomas and Schrittwieser, Julian and Antonoglou, Ioannis and Lai, Matthew and Guez, Arthur and Lanctot, Marc and Sifre, Laurent and Kumaran, Dharshan and Graepel, Thore and others},
	journal = {Science},
	number = {6419},
	pages = {1140--1144},
	publisher = {American Association for the Advancement of Science},
	title = {A general reinforcement learning algorithm that masters chess, shogi, and Go through self-play},
	volume = {362},
	year = {2018}}

@article{doi:10.1126/science.abj8754,
	author = {Minkyung Baek and Frank DiMaio and Ivan Anishchenko and Justas Dauparas and Sergey Ovchinnikov and Gyu Rie Lee and Jue Wang and Qian Cong and Lisa N. Kinch and R. Dustin Schaeffer and Claudia Mill{\'a}n and Hahnbeom Park and Carson Adams and Caleb R. Glassman and Andy DeGiovanni and Jose H. Pereira and Andria V. Rodrigues and Alberdina A. van Dijk and Ana C. Ebrecht and Diederik J. Opperman and Theo Sagmeister and Christoph Buhlheller and Tea Pavkov-Keller and Manoj K. Rathinaswamy and Udit Dalwadi and Calvin K. Yip and John E. Burke and K. Christopher Garcia and Nick V. Grishin and Paul D. Adams and Randy J. Read and David Baker},
	doi = {10.1126/science.abj8754},
	eprint = {https://www.science.org/doi/pdf/10.1126/science.abj8754},
	journal = {Science},
	number = {6557},
	pages = {871-876},
	title = {Accurate prediction of protein structures and interactions using a three-track neural network},
	url = {https://www.science.org/doi/abs/10.1126/science.abj8754},
	volume = {373},
	year = {2021}}

@article{lin2018all,
	author = {Lin, Xing and Rivenson, Yair and Yardimci, Nezih T and Veli, Muhammed and Luo, Yi and Jarrahi, Mona and Ozcan, Aydogan},
	journal = {Science},
	number = {6406},
	pages = {1004--1008},
	publisher = {American Association for the Advancement of Science},
	title = {All-optical machine learning using diffractive deep neural networks},
	volume = {361},
	year = {2018}}

@article{feldmann2019all,
	author = {Feldmann, Johannes and Youngblood, Nathan and Wright, C David and Bhaskaran, Harish and Pernice, Wolfram HP},
	journal = {Nature},
	number = {7755},
	pages = {208--214},
	publisher = {Nature Publishing Group UK London},
	title = {All-optical spiking neurosynaptic networks with self-learning capabilities},
	volume = {569},
	year = {2019}}

@article{ashtiani2022chip,
	author = {Ashtiani, Farshid and Geers, Alexander J and Aflatouni, Firooz},
	journal = {Nature},
	number = {7914},
	pages = {501--506},
	publisher = {Nature Publishing Group UK London},
	title = {An on-chip photonic deep neural network for image classification},
	volume = {606},
	year = {2022}}

@article{LeCun2015,
	author = {Yann LeCun and Yoshua Bengio and Geoffrey Hinton},
	doi = {10.1038/nature14539},
	journal = {Nature},
	month = {may},
	number = {7553},
	pages = {436--444},
	publisher = {Springer Science and Business Media {LLC}},
	title = {Deep learning},
	url = {https://doi.org/10.1038%2Fnature14539},
	volume = {521},
	year = 2015}

@article{shen2017deep,
	author = {Shen, Yichen and Harris, Nicholas C and Skirlo, Scott and Prabhu, Mihika and Baehr-Jones, Tom and Hochberg, Michael and Sun, Xin and Zhao, Shijie and Larochelle, Hugo and Englund, Dirk and others},
	journal = {Nature photonics},
	number = {7},
	pages = {441--446},
	publisher = {Nature Publishing Group UK London},
	title = {Deep learning with coherent nanophotonic circuits},
	volume = {11},
	year = {2017}}

@article{wright2022deep,
	author = {Wright, Logan G and Onodera, Tatsuhiro and Stein, Martin M and Wang, Tianyu and Schachter, Darren T and Hu, Zoey and McMahon, Peter L},
	journal = {Nature},
	number = {7894},
	pages = {549--555},
	publisher = {Nature Publishing Group UK London},
	title = {Deep physical neural networks trained with backpropagation},
	volume = {601},
	year = {2022}}

@article{Gallicchio2017,
	author = {Claudio Gallicchio and Alessio Micheli and Luca Pedrelli},
	doi = {10.1016/j.neucom.2016.12.089},
	journal = {Neurocomputing},
	pages = {87--99},
	publisher = {Elsevier {BV}},
	title = {Deep reservoir computing: A critical experimental analysis},
	url = {https://doi.org/10.1016/j.neucom.2016.12.089},
	volume = {268},
	year = {2017}}

@article{lin2022deep,
	author = {Lin, Bao-De and Shen, Yi-Wei and Tang, Jia-Yan and Yu, Jingyi and He, Xuming and Wang, Cheng},
	journal = {IEEE Journal of Selected Topics in Quantum Electronics},
	number = {6: Photonic Signal Processing},
	pages = {1--8},
	publisher = {IEEE},
	title = {Deep time-delay reservoir computing with cascading injection-locked lasers},
	volume = {29},
	year = {2022}}

@article{huang2006extreme,
	author = {Huang, Guang-Bin and Zhu, Qin-Yu and Siew, Chee-Kheong},
	journal = {Neurocomputing},
	number = {1-3},
	pages = {489--501},
	publisher = {Elsevier},
	title = {Extreme learning machine: theory and applications},
	volume = {70},
	year = {2006}}

@article{ard2022five,
	author = {Sandeep Ravidran},
	journal = {Nature},
	pages = {864--866},
	title = {Five ways deep learning has transformed image analysis},
	volume = {609},
	year = {2022}}

@article{Jaeger04,
	author = {Herbert Jaeger and Harald Haas},
	doi = {10.1126/science.1091277},
	journal = {Science},
	number = {5667},
	pages = {78--80},
	publisher = {American Association for the Advancement of Science ({AAAS})},
	title = {Harnessing Nonlinearity: Predicting Chaotic Systems and Saving Energy in Wireless Communication},
	url = {https://doi.org/10.1126/science.1091277},
	volume = {304},
	year = {2004}}

@article{jumper2021highly,
	author = {Jumper, John and Evans, Richard and Pritzel, Alexander and Green, Tim and Figurnov, Michael and Ronneberger, Olaf and Tunyasuvunakool, Kathryn and Bates, Russ and {\v{Z}}{\'\i}dek, Augustin and Potapenko, Anna and others},
	journal = {Nature},
	number = {7873},
	pages = {583--589},
	publisher = {Nature Publishing Group UK London},
	title = {Highly accurate protein structure prediction with AlphaFold},
	volume = {596},
	year = {2021}}

@article{vinckier2015high,
	author = {Vinckier, Quentin and Duport, Fran{\c{c}}ois and Smerieri, Anteo and Vandoorne, Kristof and Bienstman, Peter and Haelterman, Marc and Massar, Serge},
	journal = {Optica},
	number = {5},
	pages = {438--446},
	publisher = {Optical Society of America},
	title = {High-performance photonic reservoir computer based on a coherently driven passive cavity},
	volume = {2},
	year = {2015}}

@article{Larger17,
	author = {Larger, Laurent and Bayl{\'o}n-Fuentes, Antonio and Martinenghi, Romain and Udaltsov, Vladimir S and Chembo, Yanne K and Jacquot, Maxime},
	journal = {Physical Review X},
	number = {1},
	pages = {011015},
	publisher = {APS},
	title = {High-speed photonic reservoir computing using a time-delay-based architecture: Million words per second classification},
	volume = {7},
	year = {2017}}

@article{Liutkus2014,
	author = {Liutkus, Antoine and Martina, David and Popoff, S{\'e}bastien and Chardon, Gilles and Katz, Ori and Lerosey, Geoffroy and Gigan, Sylvain and Daudet, Laurent and Carron, Igor},
	journal = {Scientific Reports},
	pages = {5552},
	publisher = {Nature Publishing Group UK London},
	title = {Imaging With Nature: Compressive Imaging Using a Multiply Scattering Medium},
	volume = {4},
	year = {2014}}

@article{freiberger2019improving,
	author = {Freiberger, Matthias and Sackesyn, Stijn and Ma, Chonghuai and Katumba, Andrew and Bienstman, Peter and Dambre, Joni},
	journal = {IEEE Journal of Selected Topics in Quantum Electronics},
	number = {1},
	pages = {1--11},
	publisher = {IEEE},
	title = {Improving time series recognition and prediction with networks and ensembles of passive photonic reservoirs},
	volume = {26},
	year = {2019}}

@ARTICLE{Freiberger2019Training,
  author={Freiberger, Matthias and Katumba, Andrew and Bienstman, Peter and Dambre, Joni},
  journal={IEEE Transactions on Neural Networks and Learning Systems}, 
  title={Training Passive Photonic Reservoirs With Integrated Optical Readout}, 
  year={2019},
  volume={30},
  number={7},
  pages={1943-1953},
  doi={10.1109/TNNLS.2018.2874571}}

@article{wetzstein2020inference,
	author = {Wetzstein, Gordon and Ozcan, Aydogan and Gigan, Sylvain and Fan, Shanhui and Englund, Dirk and Solja{\v{c}}i{\'c}, Marin and Denz, Cornelia and Miller, David AB and Psaltis, Demetri},
	journal = {Nature},
	number = {7836},
	pages = {39--47},
	publisher = {Nature Publishing Group UK London},
	title = {Inference in artificial intelligence with deep optics and photonics},
	volume = {588},
	year = {2020}}

@article{Appeltant11,
	author = {Appeltant, Lennert and Soriano, Miguel Cornelles and Van der Sande, Guy and Danckaert, Jan and Massar, Serge and Dambre, Joni and Schrauwen, Benjamin and Mirasso, Claudio R and Fischer, Ingo},
	journal = {Nature communications},
	number = {1},
	pages = {1--6},
	publisher = {Nature Publishing Group},
	title = {Information processing using a single dynamical node as complex system},
	volume = {2},
	year = {2011}}

@article{zhou2021large,
	author = {Zhou, Tiankuang and Lin, Xing and Wu, Jiamin and Chen, Yitong and Xie, Hao and Li, Yipeng and Fan, Jingtao and Wu, Huaqiang and Fang, Lu and Dai, Qionghai},
	journal = {Nature Photonics},
	number = {5},
	pages = {367--373},
	publisher = {Nature Publishing Group UK London},
	title = {Large-scale neuromorphic optoelectronic computing with a reconfigurable diffractive processing unit},
	volume = {15},
	year = {2021}}

@article{hamerly2019large,
	author = {Hamerly, Ryan and Bernstein, Liane and Sludds, Alexander and Solja{\v{c}}i{\'c}, Marin and Englund, Dirk},
	journal = {Physical Review X},
	number = {2},
	pages = {021032},
	publisher = {APS},
	title = {Large-scale optical neural networks based on photoelectric multiplication},
	volume = {9},
	year = {2019}}

@article{Rafayelyan2020,
	author = {Rafayelyan, Mushegh and Dong, Jonathan and Tan, Yongqi and Krzakala, Florent and Gigan, Sylvain},
	journal = {Physical Review X},
	number = {4},
	pages = {041037},
	publisher = {APS},
	title = {Large-scale optical reservoir computing for spatiotemporal chaotic systems prediction},
	volume = {10},
	year = {2020}}

@article{Bengio2009,
	author = {Y. Bengio},
	doi = {10.1561/2200000006},
	journal = {Foundations and Trends in Machine Learning},
	number = {1},
	pages = {1--127},
	publisher = {Now Publishers},
	title = {Learning Deep Architectures for {AI}},
	url = {https://doi.org/10.1561%2F2200000006},
	volume = {2},
	year = 2009}

@article{Weiss95,
	author = {Carl-Otto Weiss and Udo H\"{u}bner and Neal Broadus Abraham and Dingyuan Tang},
	doi = {10.1016/1350-4495(94)00088-3},
	journal = {Infrared Physics {\&} Technology},
	number = {1},
	pages = {489--512},
	publisher = {Elsevier {BV}},
	title = {Lorenz-like chaos in {NH}3-{FIR} lasers},
	url = {https://doi.org/10.1016/1350-4495(94)00088-3},
	volume = {36},
	year = {1995}}

@article{feldmann2021parallel,
	author = {Feldmann, Johannes and Youngblood, Nathan and Karpov, Maxim and Gehring, Helge and Li, Xuan and Stappers, Maik and Le Gallo, Manuel and Fu, Xin and Lukashchuk, Anton and Raja, Arslan Sajid and others},
	journal = {Nature},
	number = {7840},
	pages = {52--58},
	publisher = {Nature Publishing Group UK London},
	title = {Parallel convolutional processing using an integrated photonic tensor core},
	volume = {589},
	year = {2021}}

@article{brunner2013parallel,
	author = {Brunner, Daniel and Soriano, Miguel C and Mirasso, Claudio R and Fischer, Ingo},
	journal = {Nature communications},
	number = {1},
	pages = {1364},
	publisher = {Nature Publishing Group UK London},
	title = {Parallel photonic information processing at gigabyte per second data rates using transient states},
	volume = {4},
	year = {2013}}

@article{triefenbach2010phoneme,
	author = {Triefenbach, Fabian and Jalalvand, Azarakhsh and Schrauwen, Benjamin and Martens, Jean-Pierre},
	journal = {Advances in neural information processing systems},
	title = {Phoneme recognition with large hierarchical reservoirs},
	volume = {23},
	year = {2010}}

@article{Butschek2022,
	author = {Lorenz Butschek and Akram Akrout and Evangelia Dimitriadou and Alessandro Lupo and Marc Haelterman and Serge Massar},
	doi = {10.1364/ol.451087},
	journal = {Optics Letters},
	month = {feb},
	number = {4},
	pages = {782},
	publisher = {Optica Publishing Group},
	title = {Photonic reservoir computer based on frequency multiplexing},
	url = {https://doi.org/10.1364%2Fol.451087},
	volume = {47},
	year = 2022}

@article{Lupo2021,
author = {Alessandro Lupo and Lorenz Butschek and Serge Massar},
journal = {Opt. Express},
number = {18},
pages = {28257--28276},
publisher = {Optica Publishing Group},
title = {Photonic extreme learning machine based on frequency multiplexing},
volume = {29},
month = {Aug},
year = {2021},
url = {https://opg.optica.org/oe/abstract.cfm?URI=oe-29-18-28257},
doi = {10.1364/OE.433535},
}

@article{jha2022photonic,
	author = {Jha, Aashu and Huang, Chaoran and Peng, Hsuan-Tung and Shastri, Bhavin and Prucnal, Paul R},
	journal = {Journal of Lightwave Technology},
	number = {9},
	pages = {2901--2914},
	publisher = {IEEE},
	title = {Photonic spiking neural networks and graphene-on-silicon spiking neurons},
	volume = {40},
	year = {2022}}

@article{Nakajima2022,
	author = {Mitsumasa Nakajima and Katsuma Inoue and Kenji Tanaka and Yasuo Kuniyoshi and Toshikazu Hashimoto and Kohei Nakajima},
	doi = {10.1038/s41467-022-35216-2},
	journal = {Nature Communications},
	number = {1},
	publisher = {Springer Science and Business Media {LLC}},
	title = {Physical deep learning with biologically inspired training method:~gradient-free approach for physical hardware},
	url = {https://doi.org/10.1038/s41467-022-35216-2},
	volume = {13},
	year = {2022}}

@article{markovic2020physics,
	author = {Markovi{\'c}, Danijela and Mizrahi, Alice and Querlioz, Damien and Grollier, Julie},
	journal = {Nature Reviews Physics},
	number = {9},
	pages = {499--510},
	publisher = {Nature Publishing Group UK London},
	title = {Physics for neuromorphic computing},
	volume = {2},
	year = {2020}}

@inproceedings{SaadeRandomProj2016,
	author = {Saade, A. and Caltagirone, F. and Carron, I. and Daudet, L. and Dr{\'e}meau, A. and Gigan, S. and Krzakala, F.},
	booktitle = {2016 IEEE International Conference on Acoustics, Speech and Signal Processing (ICASSP)},
	doi = {10.1109/ICASSP.2016.7472872},
	pages = {6215-6219},
	title = {Random projections through multiple optical scattering: Approximating Kernels at the speed of light},
	year = {2016}}

@article{Tanaka2019,
	author = {Gouhei Tanaka and Toshiyuki Yamane and Jean Benoit H{\'{e}}roux and Ryosho Nakane and Naoki Kanazawa and Seiji Takeda and Hidetoshi Numata and Daiju Nakano and Akira Hirose},
	doi = {10.1016/j.neunet.2019.03.005},
	journal = {Neural Networks},
	month = {jul},
	pages = {100--123},
	publisher = {Elsevier {BV}},
	title = {Recent advances in physical reservoir computing: A review},
	url = {https://doi.org/10.1016%2Fj.neunet.2019.03.005},
	volume = {115},
	year = 2019}

@article{nakajima2021scalable,
	author = {Nakajima, Mitsumasa and Tanaka, Kenji and Hashimoto, Toshikazu},
	journal = {Communications Physics},
	number = {1},
	pages = {20},
	publisher = {Nature Publishing Group UK London},
	title = {Scalable reservoir computing on coherent linear photonic processor},
	volume = {4},
	year = {2021}}

@article{hansen2006cma,
	author = {Hansen, Nikolaus},
	journal = {Towards a new evolutionary computation: Advances in the estimation of distribution algorithms},
	pages = {75--102},
	publisher = {Springer},
	title = {The CMA evolution strategy: a comparing review},
	year = {2006}}

@inproceedings{Kassa2018,
	author = {Wosen Kassa and Evangelia Dimitriadou and Marc Haelterman and Serge Massar and Erwin Bente},
	booktitle = {Neuro-inspired Photonic Computing},
	doi = {10.1117/12.2306176},
	editor = {Marc Sciamanna and Peter Bienstman},
	month = may,
	publisher = {{SPIE}},
	title = {Towards integrated parallel photonic reservoir computing based on frequency multiplexing},
	url = {https://doi.org/10.1117/12.2306176},
	year = {2018}}

@article{Sunada2020,
	author = {Sunada, Satoshi and Kanno, Kazutaka and Uchida, Atsushi},
	journal = {Optics Express},
	number = {21},
	pages = {30349--30361},
	publisher = {Optical Society of America},
	title = {Using multidimensional speckle dynamics for high-speed, large-scale, parallel photonic computing},
	volume = {28},
	year = {2020}}

@misc{chatgpt,
  title = {{OpenAI}},
  howpublished = {\url{https://openai.com/about}},
  note = {Accessed: March 2023}
}

@misc{bing,
  title = {{Bing Chat}},
  howpublished = {\url{https://www.bing.com}},
  note = {Accessed: March 2023}
}

\end{document}